\documentclass[12pt]{iopart}
\usepackage{iopams}
\usepackage{harvard}
\usepackage{color}
\usepackage{url}

\expandafter\let\csname equation*\endcsname\relax
\expandafter\let\csname endequation*\endcsname\relax
\usepackage{amsmath}
\usepackage{graphicx}
\usepackage[usenames,dvipsnames]{xcolor}
\usepackage{multirow}

\newcommand{\cm}{cm$^{-1}$}

\begin{document}

\title[The {\it ab initio} calculation of spectra of open shell diatomic
molecules]{The {\it ab initio} calculation of spectra of open shell diatomic
molecules}

\author{Jonathan Tennyson$^{1,2}$, Lorenzo Lodi$^{1}$, Laura K. McKemmish$^{1}$ and Sergei N Yurchenko$^{1}$}

\address{$^1$Department of Physics \& Astronomy, University College London,
Gower St.,
London, WC1E 6BT, UK}

\ead{$^2$j.tennyson@ucl.ac.uk}

\begin{abstract}
The spectra (rotational, rotation-vibrational or electronic) of diatomic molecules
due to transitions involving only closed-shell ($^1\Sigma$) electronic states
follow very regular, simple patterns and their theoretical analysis is usually
straightforward. On the other hand,
open-shell electronic states
lead to more complicated spectral patterns and, moreover,
often appear as a manifold of closely lying electronic states,
leading to perturbations with even larger complexity.
This is especially true when at least one of the atoms is a
transition metal.
Traditionally these complex cases have been analysed using
approaches based on perturbation
theory, with semi-empirical parameters determined
by fitting to spectral data.

Recently the needs of two rather diverse scientific areas  have driven
the demand for improved theoretical models of open-shell diatomic systems
based on an \emph{ab initio} approach;
these areas are ultracold
chemistry and
the astrophysics of ``cool'' stars, brown dwarfs and most recently
extrasolar planets.
However, the
complex electronic structure of these molecules combined with
the accuracy requirements
of high-resolution spectroscopy render such an approach particularly challenging.
This review  describes recent progress in developing  methods for
directly solving the effective Schr\"odinger equation for open-shell diatomic
molecules, with a focus on molecules containing a transtion metal. It considers four aspects
of the problem:
\begin{enumerate}
\item The electronic structure problem;
\item Non-perturbative treatments of the curve couplings;
\item The solution of the nuclear motion Schr{\"o}dinger equation;
\item The generation of accurate electric dipole transition intensities.
\end{enumerate}
Examples of applications are used to illustrate these issues.
\end{abstract}

\maketitle

\section{Introduction}

New and powerful experimental techniques to produce and study ultra-cold
molecules (below 1~mK) are becoming increasingly available as recently reviewed by
\citeasnoun{15HaPiLi}. These techniques include combining two ultra-cold atoms
\cite{03BeFeGu.RbCs,06JuPeKi.NaH,12UlDeRe,15PaZaSt.ai}, direct laser cooling
\cite{10ShBaDe,15GlPrEn},  deceleration techniques \cite{11HoMoMe}, 
cooling \cite{12StHuYe},
magneto-optical trapping \cite{13HuYeSt,14BaMcNo.diatom} and buffer-gas cooling
\cite{12HuLuDo}.  This has stimulated new interest in the detailed spectroscopy
of open shell diatomic systems \cite{04Dixxxx.CaH,12QuJixx}. Similarly, but at
significantly higher
temperatures, the study of molecules in the atmospheres of brown
dwarfs and cool stars requires comprehensive datasets of spectral
absorptions by a range of diatomic species whose vibronic transitions
absorb near the peak flux of the object under study
\cite{13RaReAl.NaHAlH}. Similar species are also thought to be
important constituents of the atmospheres of hot exoplanets
\cite{14AlMoPe.exo,14StBeSe.exo}, although there are as yet no
confirmed observations of such spectra from a specific molecule. The
spectra of these species also provide important diagnostics for
those interested in probing hot environments. Examples include fusion
plasmas \cite{98DuStSu.BeH,07BrPoBo.diatom,14BrStNi.BeH}, combustion
\cite{77LiBrx1.TiO,01GlSeKr.AlO}, laser ablation
\cite{00VaFoFo.AlO,03ZhLixx.AlO,06KoSexx.YO}, laser-induced plasmas
\cite{14BaMoLe.AlO,14SuPa.AlO} and discharges
\cite{09VaAsCr.FeH,13ChYeWa.C2}.  Other uses include atmospheric
studies \cite{65Joxxxx.AlO,81GoKoxx.AlO,96KnPiMu.AlO}, measuring the
magnetic fields in cool stars and brown dwarfs
\cite{99JoVaKo.TiO,06ReiBas.CrH,10ShReWe.FeH}, monitoring the state
distribution of reaction products \cite{04SoChHs.CaH,05ChChHs.CaH} and
products of chemical vapour deposition \cite{00NoKoMi.SiH}. The issues discussed in this review are relevant also to
calculations of collisions of open-shell atoms, though there are additional problems in these systems not covered in the current review.

Theoretically the study of rotationally and vibrationally resolved
electronic (rovibronic) spectra of diatomics needs to consider
couplings due to several sources of angular momentum such as
electronic spin, electron orbital angular momomentum and molecular
rotational motion.  To these can be added the hyperfine effects due to
nuclear spin angular momentum and the possible effects of external
fields. The various couplings are complicated compared to their atomic
equivalents by the presence of the nuclear axis.  These problems have
been considered over many years and have been the subject of a number
of books
\cite{Herzberg.book,69Ko.diatom,75Mi.diatom,86LeFexx,Lefebvre-Brion-Field.book,10BrCa.diatom}.
However the approach encapsulated in these works involves the
extensive use of perturbation theory to treat the various angular
momentum couplings and, often, other aspects of the problem too.
Whereas such approaches have proved effective for parameterising
experimental data, they are problematic when the couplings are strong;
in such difficult cases the  
couplings may be 
ascribed to the wrong physical mechanism \cite{jt639}.
Even vibronic
assignment of experimental spectra using these approaches remains questionable \cite{jt618}.
Furthermore, use of perturbation theory to treat the couplings is less
useful for formulating predictive, first principles studies.

Particularly challenging is the treatment of diatomic molecules
containing at least one transition metal atom. These systems often
have several low-lying electronic states, which is why they are
important astronomically, and have a complicated electronic
structure. We are
participating in a project called ExoMol which is aimed at computing
line lists for all molecular species important in hot astronomical
atmospheres, such as those found around cool stars, brown dwarfs and
exoplanets \cite{jt528}.  As part of this project it was
necessary to consider extensive rovibronic spectra of a variety of
closed shell \cite{jt563,jt583,jt590,jt605,jt615} and open shell
\cite{jt529,jt598,jt599,jt618,jtAlH} diatomics. Studies on some of the simpler systems
have used perturbation theory to include the effects of fine structure 
 \cite{jt529,jtAlH}; similar methods have been used by other
groups studying diatomics composed of light atoms
\cite{13BrBeScBa.C2,13HiWaRa.MgH,14RaBrWe.CP,14BrRsWe.CN,14BrBeWe.NH,16BrBeWe.OH}.
However, the complexity of open shell transition metal diatomics
(often combined with a shortage of experimental data) means
that their spectra may not be well-treated by this approach.
This has led us to
develop a procedure for the direct solution of the Schr\"odinger
equation augmented with coupling terms, such as spin-orbit coupling,
which arise from the inclusion of relativistic effects in the problem
\cite{jt589,jt599}. It is these procedures, and related work performed
by other groups, that concern us here. Early work in this
direction was performed by Marian  \cite{95Marian.NiH,01Marian.diatom} who
calls this a matrix procedure because, as we shall see, it is based on the
construction and diagonalisation of matrices involving the nuclear motion
wavefunctions of several electronic states to give energy levels and
associated fully coupled wavefunctions.

There are well developed programs for solving the uncoupled diatomic
nuclear motion problem such as \textsc{LEVEL} \cite{lr07}.  Although there are a
growing number of studies treating coupled electronic states
\cite{95CaLeMa.diatom,02TaFeZa.NaRb,03BeFeGu.RbCs,05MeZa.LiAr,08HuTiJu.diatom,10ZhSaDa.diatom,13GoAbHa.diatom,14BrRsWe.CN},
there appears to be a lack of general programs for the solving the
matrix procedure. We note \textsc{BOUND} by \citeasnoun{94Hutson} can be
used in this fashion. Recently we have developed a program, \textsc{DUO},
to solve the general, coupled-states diatomic nuclear motion problem
\cite{jt609}.

Modern electronic structure theory is generally
able to reproduce the basic properties of the potential
energy surfaces of open shell diatomics made
up by main group, light atoms (H to Ar).
Generally speaking because of inevitable computational limitations
the accuracy obtainable by electronic structure theory decreases
as the number of electrons in the molecule increases. For example, the
excitation energies, vibrational frequencies and bond lengths of
complicated open shell systems such as C$_2$ are well reproduced by
theory \cite{07ScBaxx.C2,11SchmBa,09NaJoPaRe.C2}.

The situation is rather worse when we consider transition metal diatomics.
In these cases not only does one  have to deal with an increased number of
electrons but also the density of electronic states is higher, leading to
more frequent perturbations.
In this review, we concentrate on these cases as they represent the most challenging situations,
specifically from the perspective of electronic structure theory.

\section{Overall Method}

When the diatomic molecule is not in a $^1\Sigma^\pm$ electronic state
interactions between
the various terms are present and one cannot treat each electronic
state in isolation.

The procedure followed starts by making the Born-Oppenheimer
approximation and separating the electronic and nuclear motion,
although it is possible to partially correct for this
approximation later in the calculation. Consideration of relativistic
and other angular momentum effects leads to couplings which are considered
in detail in the next section. The resulting nuclear motion Hamiltonian
appears quite complicated but is amenable to direct variational solution
\cite{jt618}. However, the calculation of the necessary potential
energy curves (PECs) and angular momentum coupling curves remains
challenging.

In particular, as discussed in more detail in section \ref{sec:formal},
there are three main sources of angular momentum: electronic orbital
angular momentum $\hat{\mathbf{L}}$, electronic spin angular momentum
$\hat{\mathbf{S}}$, and nuclear rotation angular momentum $\hat{\mathbf{R}}$;
in turn, these lead to a multitude of interactions
(couplings) between different electronic states. 
These angular momenta should be coupled together to form the total angular momentum
$\hat{\mathbf{J}}$, which is the only conserved angular momentum in the system.
For three angular momentum operators there are in principle 3!=6
possible ordering, and these orderings are generally denoted Hund's
cases and labelled with the letters
$(a), (b), (c), (d), (e), $ and $(e')$
\cite{Stepanov1974,10BrCa.diatom,Nikitin1994}.
For perturbative treatments
the order in which these couplings are performed matters. In
treatments which aim at a full solution of the problem they just
represent how the angular basis set is constructed and, given a
complete basis, the results should be independent of this choice.
Formal issues arising from the open shell diatomic problem have
recently been discussed in detail by \citeasnoun{15Schwenke.diatom},
who like us and \citeasnoun{95Marian.NiH}, favours the use of a Hund's
case $(a)$ representation of the problem. In this scheme
the problem is represented in terms of the angular momenta $J$ and $R$,
and the projections along the molecule's internuclear axis,
$\Lambda$, $\Sigma$ and $\Omega$, see Table~\ref{tab:ang.mom}.
Of these only $J$ and the parity are actually conserved
quantities.

Dipole moment curves (DMCs), both diagonal and
between different states, are required to give transition probabilities.
The accurate calculation of these various curves requires solution of
electronic structure problems which remain challenging in the presence
of a transition metal atom. These are discussed in section \ref{sec:dipoles}.

\begin{table}
\begin{center}
\caption{Summary of the notation used for angular momenta in diatomic molecules after \protect\citeasnoun{86LeFexx}. \label{tab:ang.mom}}
\begin{tabular}{llll}
\hline \hline
\multicolumn{1}{c}{Angular momentum}  & \multicolumn{1}{c}{Operator} &
\multicolumn{1}{c}{Projection$^a$} & \multicolumn{1}{c}{Commutation$^b$}\\
\hline
electronic orbital &  $\hat{\mathbf{L}} = \sum_i \hat{\mathbf{l}}_i$  & $\Lambda$ &  normal  \\
electron spin          &  $\hat{\mathbf{S}} = \sum_i \hat{\mathbf{s}}_i$  & $\Sigma$  &     normal  \\
total electronic   &  $\hat{\mathbf{J}}_a = \hat{\mathbf{L}} + \hat{\mathbf{S}}$  & $\Omega = \Lambda + \Sigma$  & normal  \\
\mbox{}\\
nuclear rotational &  $\hat{\mathbf{R}}$                                      & 0  &            anomalous  \\
nuclear spin           &  $\hat{\mathbf{I}} = \hat{\mathbf{I}}_a + \hat{\mathbf{I}}_b$ & $\Omega_I$ &    normal  \\
\mbox{}\\
nuclear rotational plus electronic orbital  & $\hat{\mathbf{N}} =  \hat{\mathbf{R}} + \hat{\mathbf{L}} $ & $\Lambda$ &   anomalous   \\                
nuclear rotational plus electron spin       & $\hat{\mathbf{O}} =  \hat{\mathbf{R}} + \hat{\mathbf{S}} $ & $\Sigma$  &   anomalous \\                
nuclear rotational plus total electronic    & $\hat{\mathbf{J}} =  \hat{\mathbf{R}} + \hat{\mathbf{L}} + \hat{\mathbf{S}}$ &  $\Omega$ &    anomalous\\
`total' angular momentum                    & $\mathbf{F} =  \hat{\mathbf{R}} + \hat{\mathbf{L}} + \hat{\mathbf{S}} + \hat{\mathbf{I}}$ &  $\Omega_F$ &  anomalous\\
\hline \hline
\end{tabular}
\end{center}
\mbox{}\\
$^a$ Label used for the $z$ axis (i.e., the internuclear vector).\\
$^b$ Whether the components of the given  angular momentum
along the body-fixed axes obey normal  ($[A_x,A_y] = i \hbar A_z$) or anomalous  ($[A_x,A_y] = -i \hbar A_z$) commutation
relations.\\
\end{table}

\section{Formal considerations}\label{sec:formal}
\subsection{Body-fixed Hamiltonian for diatomics}\label{sec:body-fixed}
The non-relativistic Hamiltonian in the absence of external fields
of a diatomic molecule with nuclei $A$ and $B$ and $n$ electrons is given in
atomic units
by:
\begin{equation}\begin{split}\label{fullH}
  \hat{H} &= \hat{T}_{\rm N} + \hat{T}_{\rm e} + V_{\rm NN}+ V_{\rm ee} + V_{\rm Ne} \\
 &= -\frac{\hbar^2}{2 M_A} \vec{\nabla}_A^2 -\frac{\hbar^2}{2 M_B}\vec{\nabla}_B^2
+\frac{\hbar^2}{2 m_e} \sum_{i=1}^n \vec{\nabla}_i^2 \\
 & \ +\frac{1}{4 \pi \varepsilon_0} \left\{\frac{Z_A Z_B}{|\mathbf{R}_A -
\mathbf{R}_B|} + \sum_{i<j}^n \frac{q^2}{|\mathbf{r}_i - \mathbf{r}_j|} -
\sum_{i=1}^n \left[ \frac{Z_A q}{|\mathbf{r}_i - \mathbf{R}_A|}
+\frac{Z_B q}{|\mathbf{r}_i - \mathbf{R}_B|} \right]\right\}
\end{split}
\end{equation}
where nucleus $I$ ($I=A,B$) has mass $M_I$, charge $Z_I$ and is represented by
coordinate  $\mathbf{R}_I$; $q$ is the elementary positive charge;
the electrons are represented by coordinates $\mathbf{r}_i$.
We will call the Cartesian coordinates defining the Hamiltonian above
\emph{laboratory-fixed}.

The Hamiltonian (\ref{fullH}) has the following exact symmetries \cite{jt475}:
\begin{enumerate}
\item[i)]   It is invariant to rigid translations of all coordinates.
\item[ii)]  It is invariant to rigid rotation of all coordinates.
\item[iii)] It is invariant to inversion of all coordinates.
\item[iv)]  It is invariant to permutation of any two electronic coordinates.
\item[v)]   If the two nuclei are identical, it is invariant by permutation of
the nuclear coordinates.
\end{enumerate}
Symmetry {i)} implies that the centre-of-mass motion of the system can be
separated
out. Symmetry {ii)} that the sum of orbital and rotational angular
momenta, $\hat{\mathbf{N}}$, is a conserved quantity; furthermore
the state is $N(N+1)$ times degenerate and one may refer to one specific
degenerate component by specifying
the projection of $\hat{\mathbf{N}}$ along the laboratory $Z$ axis, $M_N$. Symmetry
{iii)} indicates that parity $\tau = \pm 1$ is a
conserved quantity and can be used as a quantum number, for which the label $e$ or $f$
are generally used \cite{75BrHoHu.diatom}.
Symmetry {iv)} merely indicates that we can always satisfy the Pauli
principle (i.e., we can always choose a solution antisymmetric with respect to
exchange of any two electron coordinates). Similarly for symmetry {v)}, if the two nuclei are
identical, we can always
find a solution which is antisymmetric or symmetric with respect to nuclear
interchange; this gives rise to the well-known ortho and para states of
homonuclear diatomics \cite{98BuJexx.method}.

Hamiltonian~(\ref{fullH}) is non-relativistic so commutes with the electron and
nuclear spin operators $\hat{\mathbf{S}}$ and $\hat{\mathbf{I}}$ as well
as their projections along the laboratory $Z$ axis, so that a wave function has
a degree of degeneracy equal to $N(N+1) S(S+1) I(I+1)$.
One can refer to specific degenerate components by specifying the quantum
numbers of the projection operators along the laboratory $Z$ axis, $M_N$, $M_S$ and $M_I$,
or alternatively
by specifying the total angular momentum $\hat{\mathbf{F}} = \hat{\mathbf{N}} + \hat{\mathbf{S}}
+\hat{\mathbf{I}}$ and its component along the laboratory $Z$ axis, $M_F$. In summary,
Hamiltonian~(\ref{fullH}) admits the following exact quantum
numbers: $N$ and $M_N$ (overall angular momentum of all particules excluding
spin and its projection); $\tau$
(inversion parity quantum number); $S$ and $M_S$ (electron spin quantum number
and its projection); $I$ and $I_S$ (nuclear spin quantum number and its
projection). We note
that similar considerations hold for any molecule, not just diatomics.
Any other label used specifically for diatomics, such as $\Lambda$, must
therefore be considered an approximate quantum number.

Details on how to separate the translational and rotational
degrees of freedom can be found elsewhere
\cite{Kolos1963,Pack1968,Judd.book,Bunker1968,Sutcliffe2007,Islampour2015}.
It is customary for diatomics to choose body-fixed coordinates such that
the origin is at the centre of nuclear mass and the body-fixed $z$ axis always points
along the internuclear vector. If one deals with the rotational symmetry by
introducing new,
rotationally-invariant
(`body-fixed') coordinates and integrating out the degrees of freedom
corresponding to
the rotational symmetry \cite{Pack1968,Hornkohl1996,Sutcliffe2007,Islampour2015},
then one obtains an effective Hamiltonian depending on two quantum numbers,
namely the total angular momentum $J$ and its projection along
the body-fixed $z$ axis, $\Omega$.
The effective Hamiltonian one obtains is (see eq.~(12) of
\citeasnoun{Bunker1968}) is:
\begin{equation}\begin{split}\label{H.eff}
  \hat{H}_\mathrm{eff} & = \hat{H}_\mathrm{e} + \hat{H}_{\rm v} + \hat{H}_{\rm R} + \hat{H}_{\mu} 
\end{split}
\end{equation}
where $\hat{H}_\mathrm{e}$ is the electronic Hamiltonian and is given by
\begin{equation}
  \hat{H}_\mathrm{e} = \hat{T}_{\rm e} + V_{\rm NN}+ V_{\rm ee} + V_{\rm Ne}
\end{equation}
and the terms are defined in  Hamiltonian (\ref{fullH}).
The body-fixed vibrational energy operator $\hat{H}_\mathrm{v}$ is
\begin{equation}\begin{split}\label{vib.H}
  \hat{H}_\mathrm{v} & = - \frac{\hbar^2}{2 \mu} \frac{1}{R} \frac{\partial^2}{\partial R^2} R
\end{split}\end{equation}
where $R$ is the internuclear separation and
\begin{equation}\label{red.mass}
\mu = \left( \frac{1}{M_A} + \frac{1}{M_B} \right)^{-1}
\end{equation}
is the reduced mass of the molecule. Note that if, as is customary, a factor $1/R$
is included in the definition of the vibrational wavefunction $\phi(R)$:
\begin{equation}
\phi(R) = u(R) / R
\end{equation}
then if we choose to work with the wavefunction $u(R)$
the body-fixed vibrational energy operator $\hat{H}_\mathrm{v}$ takes
on the simpler form
\begin{equation}\label{vib.H.simple}
\hat{H}_\mathrm{v} = - \frac{\hbar^2}{2 \mu} \frac{\partial^2}{\partial R^2}
\end{equation}
The body-fixed coordinate rotational energy operator $\hat{H}_\mathrm{R}$ is
given by
\begin{equation}\label{rot.H}
  \hat{H}_\mathrm{R} = B \, \hat{\mathbf{R}}^2 
\end{equation}
where $B=\hbar^2/(2 \mu R^2)$, and $\hat{\mathbf{R}}$ is the body-fixed rotational
angular momentum;
finally, $\hat{H}_{\mu}$ is the mass-polarisation energy given by
\begin{equation}
\hat{H}_{\mu} = -\frac{\hbar^2}{8 \mu} \sum_{i=1}^N \sum_{j=1}^n \vec{\nabla}_i \cdot
\vec{\nabla}_j
\end{equation}
The mass polarisation term is small and can safely be neglected
in all but the highest accuracy calculations \cite{Kut97}.


Effects due to couplings between the electronic
potential energy curves of $\hat{H}_v$ and of the diagonal part of $\hat{H}_R$
are generally referred to as non-adiabatic and their
inclusion is important in high-accuracy studies.
While treatment of the electronically diagonal adiabatic correction is
relatively straightforward {\it ab initio} \cite{hys86,Kut97,jt153}, inclusion of
the non-adiabatic effects is not.
The relative importance of the adiabatic versus non-adiabatic correction terms is considered by
\citeasnoun{15ReMcM3.lkm}.
There have
been some attempts to treat non-adiabatic effects in a completely {\it ab initio} fashion starting
from the classic works on H$_2$ by \citeasnoun{Kolos1963}; other early studies include those by
\citeasnoun{Herman1966}, \citeasnoun{Bunker1977} and \citeasnoun{Hutson1980}.
Work on non-Born-Oppenheimer effects for H$_2$ continues \cite{14PaKoxx.H2} and,
indeed, has reached the exquisite accuracy of  of 10$^{-4}$ cm$^{-1}$
for the vibrational fundamentals of
the main isotopologues \cite{13DiNiSa.H2}.
\citeasnoun{Schwenke2001} proposed a more general method of treating non-adiabatic effects in
molecules which he tested on H$_2$$^+$.
In practical application of these approaches, e.g. by \citeasnoun{02LeHuxx.diatom}, non-adiabatic
effects are taken into account by replacing the rotational kinetic energy expression in
Eq.~\eqref{rot.H} by
\begin{equation}\label{e:dist}
\frac{\hbar^2}{2 \mu R^2} \to \frac{\hbar^2}{2 \mu R^2} \left[ 1 + \alpha(R)\right],
\end{equation}
and the vibrational kinetic energy operator in Eq.~\eqref{vib.H.simple} by
\begin{equation}
\frac{\hbar^2}{2 \mu}\frac{\partial^2}{\partial R^2} \to \frac{\hbar^2}{2 \mu}\frac{\partial}{\partial R}  \left[ 1 + \beta(R)\right] \frac{\partial}{\partial R},
\end{equation}
where the unitless functions $\alpha(r)$ and $\beta(r)$ (sometimes called the rotational/vibrational
Born-Oppenheimer breakdown function or the rotational/vibrational $g$ factors) are obtained either
semi-empirically by fitting to experimental data from more than one isotope or
by \emph{ab initio} means \cite{Sauer1998,Bak2005}.
However, as discussed under Inverse Problems below,
\cite{04Watson} noted that experimental data alone are not generally enough to fully
determine the $\alpha$ and $\beta$ functions.

This approach of Le Roy \&\ Huang
is ultimately based on a perturbative treatment of non-adiabatic effects
and may break down in presence of avoided crossings between the potential energy curves.
In such cases it is usually preferable to explicitly treat the
off-diagonal matrix elements of the kinetic energy operator \eqref{vib.H} which
can be done by borrowing
approaches originally developed to describe scattering processes \cite{Hutson1994}.
Alternatively, but possibly equivalently, one can introduce diabatic (crossing) curves \cite{Lewis1968},
as done, for example, in a recent study on C$_2$ \cite{07KoBaSc.C2}.
To the non-relativistic laboratory-fixed Hamiltonian (\ref{fullH}) one can also
add further interaction terms, see, e.g.,
Chapter 3 of \citeasnoun{10BrCa.diatom} for a derivation and list of
interactions.
In particular, the spin-orbit Hamiltonian operator is usually essential for even
qualitative accuracy of transition metal diatomic spectra while
the spin-rotation and spin-spin operators are sometimes important in high
accuracy studies.
The coupling rules which govern the interactions between
these terms are given in Table~\ref{t:sel-rules}.

In principle, these terms can be evaluated once the spin is included into the wavefunction.
However, only the spin-orbit matrix elements are routinely computed within specialised
electronic structure programs. The one-electron component of the
spin-orbit coupling usually dominates; therefore, treatment of electron
correlation is less critical than for other properties \cite{11LiMuSz.ai}.
For the other spin-terms, we in practice often use effective coupling terms for
the spin-spin and spin-rotational Hamiltonian operators and fit to experimental
energies or frequencies. These are discussed in more detail by
\citeasnoun{jt609}.

\subsection{Wavefunctions}

Within a Hund's case $(a)$ frawework, a complete basis set can be
constructed by multiplying the electronic wave
functions  by a set of functions complete for the nuclear (vibrational
and rotational) degrees of freedom, giving:
\begin{equation}\label{e:basis}
 |{\rm state}, J, \Omega, \Lambda, S, \Sigma, v \rangle  =  | {\rm state},
\Lambda, S, \Sigma  \rangle | J,\Omega,M  \rangle |{\rm state}, J, S, \Lambda, \Omega, v \rangle,
\end{equation}
where (see also table~\ref{tab:ang.mom}) $\Lambda = 0, \pm 1, \pm 2, \cdots$ is the
eigenvalue of $\hat{L}_z$; $S$ is the electron spin quantum number;
$M$ is the projection of the total angular momentum along the
laboratory axis $Z$; $v$ is the vibrational quantum number. For the
rovibronic basis set the combinations of $\Sigma$ and $\Lambda$ are
selected to satisfy $-J \le \Omega \le J$.  The label `state' is a
counter over electronic states having the same symmetry.
The $| {\rm state}, \Lambda, S, \Sigma  \rangle$  are solutions of the electronic structure problem
and can be obtained (although usually not exploiting the full $C_{\infty v}$ or
$D_{\infty h}$ point-group symmetry of a diatomic system) using
a number of electronic structure packages;
the $| J, \Omega,M \rangle$ are symmetric top eigenfunctions
which are Wigner
$\mathcal{D}$ functions \cite{Pack1968,Sutcliffe2007,Zare.book,Islampour2015}
and depend on the Euler angles ($\alpha, \beta, \gamma)$ specifying the
orientation of the body-fixed axis with respect to the space-fixed one:
\begin{equation}
| J, \Omega,M \rangle =  \sqrt{\frac{2J+1}{4 \pi}}
\mathcal{D}^J_{M,\Omega}(\alpha, \beta, \gamma=\frac{\pi}{2} )^{*}
\end{equation}
Note that in the expression above the third Euler angle, $\gamma$, has been
fixed to the value $\pi /2$ as the condition
that the body-fixed $z$ axis points from one nucleus
to the other leaves it completely unconstrained; the normalization factor
has been chosen so that that no integration over $\gamma$ should be performed.
This is the customary choice for diatomics \cite{vanVleck1951},
although other treatments of $\gamma$ are possible \cite{Judd.book,Hornkohl1996}.

Finally,  $|{\rm state}, J, S, \Lambda, \Omega, v \rangle$ are vibrational functions found as solutions
of the one-dimensional Schr{\"o}dinger equations
for the electronic potential curves $E_n(r)$ (possibly complemented
by the diagonal part of the rotational Hamiltonian, eq.~\eqref{Hr.diagonal})
and will be discussed in detail in the next section.

Note that our notation implies that the electronic wave functions
$ | {\rm state}, \Lambda, S, \Sigma \rangle$  are chosen such that:
\begin{eqnarray}
\langle  {\rm state}, \Lambda, S, \Sigma  |\hat{L}_{z}| {\rm state}, \Lambda, S,
\Sigma \rangle &=& \Lambda \label{e:<|Lz|>}\\
\hat{\sigma}_{v}(xz)  | {\rm state}, \Lambda, S, \Sigma \rangle &=&
(-1)^{s_q+\Lambda+S-\Sigma} | {\rm state}, -\Lambda, S, -\Sigma \rangle
\label{e:sigmav}
\end{eqnarray}
where $\hat{\sigma}_{v}(xz) $ is the operator corresponding to reflection
through the body-fixed $xz$-plane and where $s_q=1$ for $\Sigma^-$
electronic states and $s_q=0$ in all other cases;
commonly used quantum chemistry programs such as
\textsc{Molpro}~\cite{12WeKnKn.methods} do not use this form and therefore
couplings obtained from such programs need to be converted
\cite{jt589,15Schwenke.diatom,jt609}.

As a second and final step one computes the matrix elements of the Hamiltonian
(\ref{H.eff}) (plus, if desired, any
other additional Hamiltonian components such as spin-orbit coupling,
spin-rotation interaction etc.) in the chosen basis set and
obtains the final energies and wave functions by diagonalisation of the
Hamiltonian matrix.

The exact wave function has a definite parity with respect to
inversion of all laboratory-fixed coordinates; on the other hand the
basis functions given of Eq.~(\ref{e:basis}) do not. Laboratory-fixed
inversion is equivalent to reflection through the molecule-fixed $xz$
plane \cite{Pack1968,Roeggen1971,93Kato.methods}.  Symmetrized basis functions
having definite parity $\tau=\pm 1$ are given by
\begin{equation} \begin{split} \label{e:symmetrised}
 | {\rm state}, J, |\Omega|, |\Lambda|, S, |\Sigma|, v, \tau \rangle & =
\frac{1}{\sqrt{2}} \left( | {\rm state}, J, \Omega, \Lambda, S, \Sigma, v \rangle \right. \\
         &  \left. \quad + \tau \, (-1)^{s_q+J-S} | {\rm state}, J, -\Omega, -\Lambda, S, -\Sigma, v \rangle \right).
\end{split}\end{equation}
This parity relates to the standard linear-molecule $e$/$f$ parity labels \cite{75BrHoHu.diatom}
as follows. For Bosons ($J$ integer) $e$ states have parity given by $(-1)^J$ and $f$ states by
$(-1)^{(J+1)}$; for Fermions this becomes $(-1)^{(J-\frac{1}{2})}$ for $e$ states and
$(-1)^{(J+\frac{1}{2})}$ for $f$ states.

\subsection{Rotational couplings}

Actual calculations require matrix elements of the
effective Hamiltonian (\ref{H.eff}).
In the basis set functions given by eq.~(\ref{e:basis}),
the rotational angular momentum $\mathbf{\hat{R}}$ appearing in eq.~(\ref{rot.H})
can be expanded as (see table~\ref{tab:ang.mom})
\begin{equation}
\mathbf{\hat{R}} = \mathbf{\hat{J}}-\mathbf{\hat{L}}-\mathbf{\hat{S}}.
\end{equation}

The rotational Hamiltonian given by eq.~(\ref{rot.H}) can be written as \cite{86LeFexx}
\begin{eqnarray}\label{rot.ham}
 \hat{H}_R  & = \frac{\hbar^2}{2 \mu r^2} \left\{ (\hat{J}^{2}-\hat{J}_{z}^{2})+
(\hat{S}^{2}-\hat{S}^2_{z}) + (\hat{L}^{2}-\hat{L}^{2}_{z})  \right.
\label{e:Hr} \\
            & + (\hat{J}_{+}\hat{S}_{-}+\hat{J}_{-}\hat{S}_{+}) \label{e:Hr.js}
\\
            & - (\hat{J}_{+}\hat{L}_{-}+\hat{J}_{-}\hat{L}_{+}) \label{e:Hr.jl}
\\
            & \left. + (\hat{L}_{+}\hat{S}_{-}+\hat{L}_{-}\hat{S}_{+})
\right\} . \label{e:Hr.sl}
\end{eqnarray}
The second term in $\hat{H}_R$ (eq.~\ref{e:Hr.js}) is due to terms of the type
$\hat{J}_\pm\hat{S}_\mp$ and 
is called spin uncoupling, see, for example page 626 of \citeasnoun{10BrCa.diatom}
for a description.
This term leads, for increasing $J$, to a transition from Hund's case $(a)$ to case
$(d)$.
The third term in $\hat{H}_R$ (eq.~\ref{e:Hr.jl}) is due to terms of the
type $\hat{J}_\pm\hat{L}_\mp$ and 
is called $L$-uncoupling. This term is responsible for the $\Lambda$-doubling of
$\Pi$, $\Delta$, $\Phi$ states (i.e., the splitting of the $\tau = +$
and $\tau = -$ components of non-$\Sigma$ states). When it is the prevailing
perturbation it leads for increasing $J$ to a transition from Hund's
case $(a)$ to case $(d)$.
The fourth term in $\hat{H}_R$ (eq.~\ref{e:Hr.sl}) is due to terms of the type
$\hat{L}_\pm\hat{S}_\mp$ is a spin-electronic term
which obeys coupling rules very similar to the spin-orbit Hamiltonian and hence
experimentally it cannot generally be distinguished
from spin-orbit interactions.

Matrix elements of the rotational Hamiltonian (\ref{e:Hr}--\ref{e:Hr.sl}) in the
basis given by eq.~(\ref{e:basis}) 
can be evaluated using the rules of the angular momentum
\cite{86LeFexx}. 
The coupling rules for the rotational Hamiltonian, as well as for other
components of the total Hamiltonian~(\ref{H.eff})
(spin-orbit, spin rotation and spin-spin interactions)
are given in Table~\ref{t:sel-rules}.

\begin{table}
\begin{center}
\caption{Coupling rules for matrix elements of the body-fixed Hamiltonian given
by eq.~(\ref{H.eff}) using the (non-symmetrised) basis functions of
eq.~(\ref{e:basis}). The coupling rules $\Delta J =0$ and $g \nleftrightarrow
u$ (only for homonuclear diatomics) always hold. For the symmetrised basis
functions given by eq.~(\ref{e:symmetrised}) there is a further universal
selection rule, namely $e \leftrightarrow e$ and $f \leftrightarrow f$ are
possible but not $e \leftrightarrow f$.}
\label{t:sel-rules}
\begin{tabular}{l c c c ccccc}
\hline
\multicolumn{1}{c}{Description}                                 &    Eq.$^a$
      &  \multicolumn{5}{c}{Coupling rules} \\
                                                                &
      &  $|\Delta S|$ & $|\Delta \Sigma|$ & $|\Delta \Lambda|$ & $|\Delta
\Omega|$ & notes \\
\cline{3-7}
vibrational and nonadiabatic, $\hat{H}_\mathrm{v}$               &
(\ref{vib.H})      &  0          &     0         &      0           &   0  & b\\
rotational diagonal                                             &   (\ref{e:Hr})
      &  0          &     0         &      0           &   0               \\
S-uncoupling, $\hat{J}_\pm\hat{S}_\mp$                          &
(\ref{e:Hr.js})    &  0          &    $1$        &      0           & $1$       \\
L-uncoupling, $\hat{J}_\pm\hat{L}_\mp$                          &
(\ref{e:Hr.jl})    &  0          &     0         &      1           &   1       \\
spin-electronic, $\hat{L}_\pm\hat{S}_\mp$                       &
(\ref{e:Hr.sl})    &  0          &    $1$        &      $1$           &   0    \\
spin-orbit, $\hat{H}_{\rm SO}$                                  &
c&  $0,1$      &  $0, 1$       & 0, $1$           &   0   &d,e\\
spin-rotation, $\hat{H}_{\rm SR}$                               &
c&  0          &   1           &      0           &   1 \\
spin-spin, $\hat{H}_{\rm SS}$                                   &
c &  0,1,2      &   0,1,2       &      0,1,2       &   0   & f,g\\
\hline
\hline
\end{tabular}
\end{center}
\mbox{}\\
$^a$ Relevant equation in the text defining the operator.\\
$^b$ Matrix elements of $\hat{H}_\mathrm{v}$ between different electronic states
(having the same molecular term symbol) are considered nonadiabatic
interactions.\\
$^c$ See Chapter 3 of \citeasnoun{10BrCa.diatom}.\\
$^d$ Extra selection rule for $\Sigma$ electronic states: $\Sigma^\pm
\leftrightarrow  \Sigma^\mp$ but $\Sigma^\pm \nleftrightarrow  \Sigma^\pm$.\\
$^e$ The matrix element is zero if $\Delta S = 0$ and $\Sigma' = \Sigma" =0$. \\
$^f$ Extra selection rule for $\Sigma$ electronic states: $\Sigma^\pm
\leftrightarrow  \Sigma^\pm$ but $\Sigma^\pm \nleftrightarrow  \Sigma^\mp$;
the converse of that for $\hat{H}_{\rm SO}$), see note d.\\
$^g$ Spin-spin interaction is zero between $\Sigma$ states with $S \leq 1/2$.
\end{table}


The expectation value of the first term in $\hat{H}_R$, eq.~(\ref{e:Hr}),
for each of the basis functions (\ref{e:basis})
gives rise to the well-known rotational centrifugal potential.
Its effect amounts to an additive potential term to the
Born-Oppenheimer potential which can be written
in the following equivalent ways:
\begin{equation}\begin{split}\label{Hr.diagonal}
H_R^\mathrm{diag} & =   \langle J,\Omega,M |  \langle {\rm state}, \Lambda, S, \Sigma   |\hat{H}_R | {\rm state}, \Lambda, S, \Sigma  \rangle | J,\Omega,M  \rangle \\
                        & = B \left[ [J(J+1)-\Omega^2] + [S(S+1)-\Sigma^2] +
L_x^2 + L_y^2 -\Lambda^2 \right] \\
                        & = B \left[ J(J+1) + S(S+1) -2 (\Lambda^2 + \Lambda
\Sigma + \Sigma^2 ) + L_x^2 + L_y^2 \right]\\
                        & = B \left[ J(J+1) + S(S+1) -2 \Omega (\Lambda +
\Omega) + L_x^2 + L_y^2 \right].
\end{split}\end{equation}
For $^1\Sigma$ states the expression above reduces to the well-known
expression $ B J (J+1)$.
In the absence of $H_R^\mathrm{diag}$ the non-relativistic wavefunctions are
$2 (2S+1)$-times degenerate for $\Lambda \neq 0$
and $(2S+1)$-times for $\Lambda =0$.
Inclusion of $H_R^\mathrm{diag}$
partially splits these degenerate components according to their value of
$\Lambda$ and $\Sigma$. Specifically, for $S < \lfloor \Lambda / 2\rfloor $
each state splits by effect of $H_R^\mathrm{diag}$ in $2S+1$ components,
while for $S \geq \lfloor \Lambda / 2\rfloor $
each state splits into  $\lfloor S + \Lambda/2 + 1\rfloor $ components, where
$\lfloor x \rfloor$ indicates the floor function, i.e.
the largest integer not greater than $x$.
In {\it ab initio} treatments $L_x^2$ and $L_y^2$ are functions of $R$ obtained
by averaging the body-fixed angular momentum
operators over the electronic wave function. Essentially they amount
to a small correction to the Born-Oppenheimer potential $V(R)$ and in
experimental analyses they are not considered separately from it.
Because of the cylindrical symmetry of the electronic problem $L_x(R)
= L_y(R)$.  Note that the values of $L_x$ and $L_y$ depend on the
choice of the origin of the body-fixed axes, and for consistency with
the rest of the treatment, we recommend that they are computed with
the origin fixed at the centre of nuclear mass. Because of this
choice of the origin upon dissociation $L_x(R)$ and $L_y(R)$ grow
proportionally to $R$, and the term $B(L_x^2 +L_y^2)$ in eq.~\eqref{Hr.diagonal}
tends to a constant value as $B \propto 1/R^2$.

\section{Electronic structure calculations}

\subsection{Overview}
The above formalism is largely concerned with the couplings of the various
angular momenta and is, in principle,
exact.  
The largest source of uncertainty in computing the energy levels
arises from solution of the electronic structure problem
to give the required potential energy curves (PECs) and couplings between
electronic states, specifically the spin-orbit coupling and
L-uncoupling matrix elements. At the moment
only a few electronic structure packages, such as Dalton \cite{Dalton2016}, are capable of computing
the spin-rotation and spin-spin matrix elements
and such terms are usually dealt with empirically.

{\it Ab initio} electronic structure theory has been used to predict the
properties of many electronic states in diatomics containing one
transition metal. There are many recent examples of such calculations
\cite{06TzMaxx.Re,09KaMaxx.Re,10MiMaxx.TiO,11GoMaxx.Re,11SaMiMa.FeO,12DeAlxx.Re,12LeShxx.Re,12SaMaxx.Re,12SaMaxx2.Re,13SaMaxx.Re,14BoGoCa.Re,14DeHaAl.Re,00KaMaxx.Re,04BoMaxx.Re,05AbAlKo.Re,06MiBeBa.Re,06Raxxxx.Re,07DeCeTh.Re,07GhBoMa.Re,08KoMaxx.Re,08RiRoOr.Re,10SaPaMa.Re,11Paxxxx.Re,11ShLiSu.Re,14KaChGr,14KaGr,15ShLiWa.Re}.
There has also been investigations into using {\it ab initio}
electronic structure theory to study the even more complicated problem
of electronic states in diatomics containing two transition metals
\cite{06GaYaRe.Re,07ZhYaGa.Re,97CzReSt.Re,09GaYaWa.Re,10CaWiCi.Re,10KaKaMa.Re,11Kaxxxx.Re,12KrMoKa.Re,14CoMuxx.Re,14KaMixx.Re,15Duxxxx.Re,15Kaxxxx.Re}.
For example, the ground electronic state of the chromium dimer, Cr$_2$, is
a remarkably challenging system and is often used as a benchmark to test
new theoretical methods \cite{Muller2009,Kurashige2011,Sokolov2016}.

If one wants to predict or model to the highest possible
accuracy the rovibronic spectroscopy of transition
metal diatomics it is usually imperative to use
multireference wavefunction-based methodologies
which are able to treat both the so-called static and
dynamic parts of electron correlation, see the
review by \citeasnoun{12SzMuGi.ai} for a general discussion.
These approaches generally take as a starting point
a multireference self-consistent field (MCSCF) wave function,
very often of the complete active space (CASSCF) type \cite{Roos1980,Olsen2011},
and improve upon it using either perturbation theory or on matrix
diagonalization of an opportune, reduced-size Hamiltonian matrix.
Methods bases on perturbation theory include the popular
CASPT2 method \cite{Andersson1990,Andersson1992},
the $n$-electron valence state perturbation theory (NEVPT) method \cite{Angeli2001,Angeli2007},
as well as others \cite{Werner1996,Celani2000,Rolik2015},
while methods based on diagonalization include internally-contracted multireference
configuration interaction (icMRCI)\cite{12SzMuGi.ai} and closely
related modifications such as averaged coupled
pair functional (ACPF) \cite{Gdanitz1988} and
averaged quadratic coupled cluster (AQCC) \cite{Szalay1995,Szalay2008}.
Among electronic structure methods which are generally applicable both
to the ground as well as to excited electronic states (also for highly stretched
geometries) icMRCI-type ones are the most accurate, and in the following
sections we will deal solely with them. In the remainder of
the present section we briefly discuss some other alternatives.

Among single-reference electronic structure methods, i.e. methods which
take as a starting point a Hartree-Fock type wavefunction, the most
successful and widely applied is probably the coupled-cluster (CC) method,
\cite{Bartlett2007}, especially in its (closed-shell) CCSD(T) version
\cite{Raghavachari1989}. CC methods are generally used to compute
properties of the ground electronic state of
closed-shell systems close to equilibrium, and when one of these
three qualifications is violated complications arise.
For example, CC approaches for open-shell system do exists but are
much less developed \cite{Watts1993,Knowles1993,Knowles2000}
and may not give access to all possible values of spin or symmetries.
For highly stretched geometries CC methods are inevitably
expected to degrade very significantly in quality or even to break
down, so that complete energy curves up to dissociation cannot in general
be computed. Multireference generalization of coupled-cluster (MRCC)
are being developed but at the moment they are not in widespread use
nor they can provide better accuracy than icMRCI-type ones
\cite{Lyakh2011,Szalay2010,Rolik2014}.

Finally, a comment is warranted on density functional theory (DFT)
methods, which have
proved very successful in applications to main group species and
material science.
The DFT methodology is based on the choice of an
exchange and correlation functional, and a functional
appropriate for transition metal system has proved difficult
to obtain and, despite recent progress in this direction
\cite{Zhao2008,Cramer2009,Cohen2012}, DFT methods are currently
incapable of providing the necessary accuracy for high-resolution
electronic spectroscopy of transition metal diatomics.


\subsection{Computational considerations}
It may be thought that for such small (two atoms) systems calculation
time is not a problem and that most {\it ab initio} quantum chemistry
methods are practical; however, this is definitely not the case.  Some
methodology changes have practically no influence on calculation time, e.g.
inclusion of \emph{a posteriori} size-consistency corrections
such as the Davidson correction (+Q) \cite{Langhoff1974,12SzMuGi.ai}
and relativistic corrections using the mass-velocity and Darwin terms.
Diagonal and parallel off-diagonal finite field dipole moment
calculations increase calculation time by up to a factor of three.
Changes to the way in which the complete active space self-consistent
field (CASSCF) orbitals are optimised can cause or alleviate
convergence problems, but otherwise have little impact on calculation
time.
Increases to the basis set size have a significant effect on
calculation time; however, it is normally practical to perform
calculations for diatomic systems with basis sets with sizes
similar to the aug-cc-pVQZ basis set from the Dunning correlation-consistent
family \cite{Dunning1989,Balabanov2005}. The use of effective core potentials is a possible choice instead of all electron atomic basis sets. However, since we are mostly considering relatively light transition metals species with modest relativistic effects, we do not consider effective core potentials in detail here. Readers are invited to look at the recent review \cite{12DoCaxx.ecp} for further discussion. Basis set superposition errors is also a consideration, particularly for calculating dissociation energies. However, in practice, other sources of error are found to be dominant, particularly when large basis sets are used. Furthermore, methods of addressing basis set superposition error are well established, for example, using standard counterpoise corrections.

There are also some improvements to methodology that pose such
substantial increases to calculation time and memory requirements that
they are often not practical for calculations along the full bond
distance curve or for excited electronic states, though benchmark
calculations at a single bond length may be feasible. Inclusion of
semi-core correlation (e.g. including correlation from the $3s$ and $3p$
first-row transition metal electrons) falls into this category, especially when
the main group element is heavier (e.g. oxygen, fluorine). Also,
increases to the active space, such as including a second set of $d$
orbitals, pose often insurmountable practical problems. This is
unfortunate as both of these improvements would be
highly desirable for transition metal systems.


As we shall see, current {\it ab initio} electronic structure
calculations fail to achieve high accuracy for the spectroscopy of
transition metal diatomics for some properties, particularly
excitation energies.  Therefore, fitting of PECs and other curves to experimental results
is often used. However, these lower
accuracy calculations can be used to demonstrate the complexity and
nature of electronic states within a certain spectral
band. For example, the study by \cite{15HuHoHi.VO} of  the excited
states of VO  demonstrates the density of doublet states within 20~000
\cm{} and 30~000 \cm{}  is so high that correct modelling of small
perturbations in the spectra of the C $^4\Sigma^-$ and D $^4\Pi$ state
(origins at 17~420 \cm{} and 19~148 \cm{} respectively) is not
currently feasible {\it ab initio}.

\subsection{icMRCI Orbitals and Methodology}
icMRCI calculations are far from straightforward and depend subtly on a
number of key methodological parameters that are rarely explicitly
discussed. We recommend a much more thorough specification of the
methodological choices than is currently given; the standard brief
description such as
icMRCI/aug-cc-pVDZ is inadequate to allow reproducibility
of the calculation.

icMRCI calculations use molecular orbitals that are usually derived from
a CASSCF calculation. A CASSCF wavefunction is a linear combination of Slater
determinants which all share a set of core doubly occupied orbitals
and have variable occupation of a shared set of active orbitals. In a
CASSCF calculation, these orbitals are optimised according to some
criteria. For state-specific CASSCF, one electronic state with
particular spin and symmetry is specified; the orbitals are optimised
in order to minimize the energy of the CASSCF wavefunction of that
state. During this process, both the orbitals themselves and the
coefficients of each Slater determinant are optimised. For
state-averaged CASSCF calculations, there are multiple CASSCF
wavefunctions for each electronic state that share a single set of
core and active orbitals, but vary in the coefficients of each Slater
determinant. The orbitals and coefficients are all optimised to
minimise the sum of the energy of the CASSCF wavefunctions for each
state. It is also possible to put a weight on each state to preference
optimisation of one electronic state over others. Though the CASSCF
calculation produces energies, wavefunctions and other property
evaluations for each component electronic state, none of this is used
in the subsequent icMRCI calculation. The icMRCI calculation only uses the
set of core, active and virtual orbitals produced during the CASSCF
calculation (note the virtual orbitals are not occupied in any Slater
determinant but are formed from orthogonal components of the atomic
basis set). Only a single set of
orbitals is produced regardless of the number of states involved in
the CASSCF calculation. Thus, if one is interested in, say, the
properties of state A, the inclusion of any other states in the CASSCF
calculation will reduce the ability of the CASSCF orbitals to model
the A state. Thus, icMRCI calculations using state-specific CASSCF
orbitals should be better than icMRCI calculations using state-averaged
CASSCF orbitals; for example, the energy of the former should be lower
than the latter, and thus closer to the true answer since icMRCI
calculations are variational.  The only exceptions to this are (1)
accidental cancellation of errors can mean an inferior methodology
happens to reproduce experiment more accurately and (2) the dynamic
correlation is very strong and it happens that state-averaged CASSCF
orbitals can more accurately reproduce the dynamic correlation
effects. Note that this latter case is not expected to arise very
often and does not provide a justification for state-averaged CASSCF
calculations in general.

Despite the superiority of state-specific calculations from a
theoretical perspective, there are a number of situations where one
might require state-averaged (SA) calculations, or where their use may
be justified. 
First, it is imperative to include all lower-lying states of the same
spin and symmetry in a CASSCF calculation.
Second, currently there are no programs that allow computing spin-orbit matrix elements when the bra and ket do not originate from the
same orthonormal set of CASSCF orbitals;
in such cases it is therefore necessary to include both states
in the optimisation of these orbitals. However, this is a limitation that could be addressed in future code development, and we would strongly endorse this.

In some cases, convergence
can be improved by reducing symmetry from the linear point groups
$C_{\infty v}$ and $D_{\infty h}$, which means further low-lying
states of the same symmetry arise. Also, sometimes convergence is
improved by including more states; though various arguments can be
developed, this behaviour to our knowledge does not have a solid
theoretical basis, but CASSCF calculations are notoriously troublesome
and quirky in their convergence properties.  Finally, SA calculations
mean that all properties of a set of an electronic state can be
determined together.  This can be extremely helpful to ensuring the
correct phase and relative signs of dipole moments, spin-orbit and
electronic angular momentum matrix elements, particularly in a complex
system (in this case, the absolute value of these matrix elements can
be taken from icMRCI calculations using more accurate state-specific or
minimal state CASSCF orbitals). Alternatively, sometimes SA
calculations can be performed as a matter of convenience and reduced
human effort and calculation time.  However, we note that a single
SA-CASSCF with many electronic states will almost certainly take
longer than a single state-specific CASSCF. As we shall see, there are
many sources of inaccuracy in current {\it ab initio} electronic structure
theory calculations for transition metal diatomics and for many
applications the error associated with state-averaging may not be
important.  However, in the latter case, we must caution that
excessive consideration of other methodological choices, most notably
basis set, are not justified. The errors associated with the use of
double zeta basis set versus quadrupole zeta basis sets are found to be lower
 than the error associated with state-averaged calculations.

Our experiences \cite{jt623} show that the effect of the choice of
electronic states in the optimisation of the CASSCF orbitals can be
large. However,  we propose a
standardised notation to describe the way in which the CASSCF orbitals
for a particular icMRCI calculation are optimised. The states optimised
in the CASSCF orbitals are put in brackets, separated by commas, then
a dash followed by the words CAS are used. For example, (X,A)-CAS
indicates CASSCF orbitals optimised for the X and A electronic states.
If fractional weights are utilised, these are put before the state,
e.g. (0.5 X, 1 A)-CAS.

Another often forgotten methodological consideration is that in
internally-contracted
MRCI \cite{88WeKnxx.ai} --- implemented in the
popular package Molpro \cite{12WeKnKn.methods} ---
the number of states requested in the icMRCI calculation affects the result.
If more states
are requested, less contraction occurs; this results in a longer
calculation and lower absolute energies. We recommend that the number
of states requested in the icMRCI calculation be denoted in brackets
after the icMRCI; e.g. icMRCI(2) indicates a icMRCI calculation where 2
states were requested. More complicated notation (e.g. icMRCI(2,1,1,2)) could be
used to denote multiple states of different spin/symmetries. Rather than
explicitly prescribing this more complicated notation here, we recommend
describing the notation within the paper itself and ensuring that the
point group, symmetry and spin of states are fully specified.
Even better to ensure reproducibility of results, we encourage authors to provide sample input files to the electronic structure program used.

\subsection{Correct Identification of Electronic State}
The high density of electronic states for transition metal diatomics
and the relative high inaccuracies in excitation energies (see below)
makes the correct identification of electronic states in a calculation
imperative, particularly when comparing against experiment and
between different calculations. Particularly at higher energies, the
order of the electronic states is often incorrectly predicted by {\it ab
initio} procedures; furthermore, avoided and true crossing mean
that the nature of the electronic state can change with bond
length. Thus, careful monitoring of each electronic state in an {\it ab
initio} calculation is imperative. The most obvious way is through
monitoring the $L_z$ expectation value; however, this cannot
distinguish between two states of the same spin and symmetry (e.g. if
the calculation has converged on the wrong state). We find a much more
sensitive test for non-$\Sigma$ and non-singlet states is the diagonal
spin-orbit matrix element, which shows high stability across different
quality wavefunctions. Furthermore, this can be directly compared
to experiment as the
matrix element is equal to $S\Lambda A$ where $A$ is
the standard experimental spin-orbit splitting constant.
Diagonal dipole moments
are another possible diagnostic, but  generally only distinguishes between
states of very different nature (e.g. different valency). The energy
is somewhat useful, but must be considered with care; the
errors in excitation energies can make this a misleading
diagnostic, especially for low-level calculations.  For example, CASSCF excitation
energies may be double the true experimental excitation
energies  and higher electronic states, where incorrect state ordering
even at relatively high levels of theory are common.

\subsection{Potential energy curves}
When considering {\it ab initio} potential energy curves, we are most
concerned with the parameters of each electronic state, particularly
their energy relative to each other and to the ground state, the
vibrational frequencies, equilibrium bond lengths and the long range
behaviour. 
The choice of electronic
structure theory procedure is limited by available methodologies in
common quantum chemistry packages  and time
considerations. However, the most important limitation is the need for
converged calculations and for continuous, smooth curves. These
requirements are particularly difficult for the multireference
methods that are required for stretched bonds and excited states.
Furthermore, open-shell systems in general exhibit more complexity and
difficulties with convergence.

Of course there are an infinite number of electronic states converging
on any ionisation limit. This situation is not well treated by
standard electronic stucture methods. However high-lying PECs have
been successfully identified from negative energy scattering
calculations \cite{jt106}; under these circumstances the states are
often best characterised via their quantum defects \cite{jt196}. This
procedure, unlike more standard basis set methods, becomes more
accurate for very high-lying states as the error is often
characterised by a constant shift in the quantum defect
\cite{jt189,jt260}.  Quantum defect theory has been used to represent
the spectrum of highly excited H$_2$ molecules with outstanding
accuracy \cite{13SpJuMe.H2,14SpJuMe.H2}.

The most ambitious use to date of scattering theory for excited bound
states was by \citeasnoun{jt560}, who considered Rydberg states of
N$_2$ and mapped out in detail the many avoided crossings for
this system. Use of this procedure would be interesting for transition
metal containing molecules but would require further work to be done
on improving the representation of the so-called target wavefunction
(the wavefunction of the ionised system) used in the scattering model
before really accurate results can be obtained for these systems.

\paragraph*{Dissociation energies} are calculated from two
single-point energy calculations; therefore, very high level
calculations are possible \cite{jt549}. However, there appears to be a
relative lack of interest in this property and satisfaction with
experimental results.  This means that very high accuracy methodologies have not
been extensively explored for transition metal diatomics.

\begin{figure}
\includegraphics[width=\textwidth]{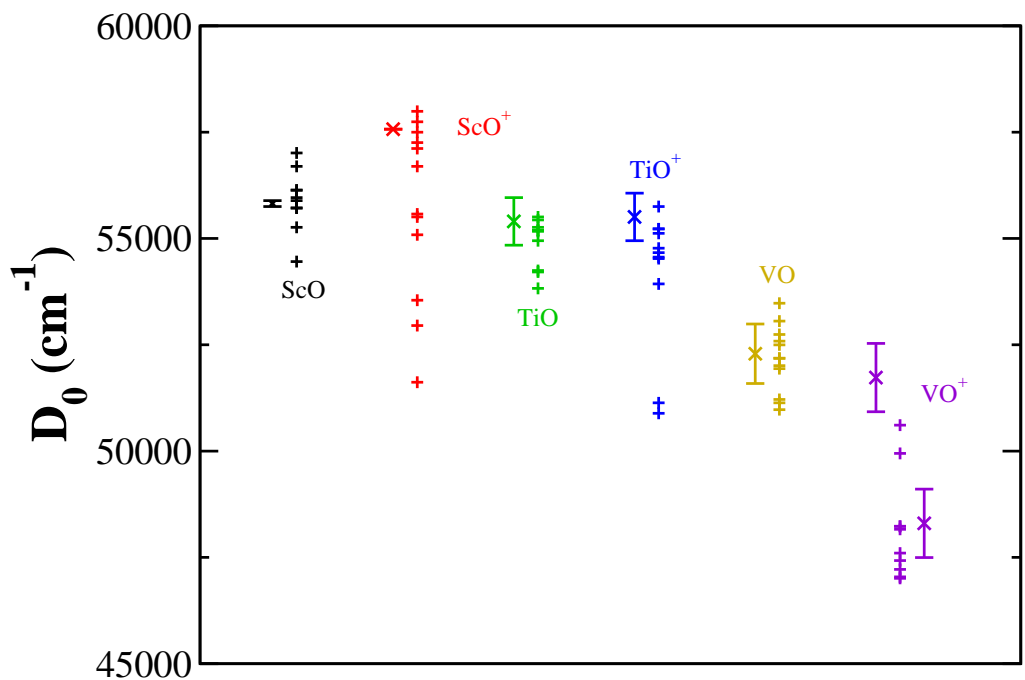}
  \caption{\label{fig:DE}  Calculated {\it ab initio} dissociation energies of
    neutral and charged transition metal oxides, using large basis
    sets and a variety of treatments of static and dynamic
    correlation compared to experiment (with error bars).
{\it Ab initio} results are taken from \protect\citeasnoun{91TiHaxx.ai}, \protect\citeasnoun{95BaMaxx.ai},
\protect\citeasnoun{01NaHiTa.ai}, \protect\citeasnoun{07MiMaxx.VO},  and
\protect\citeasnoun{10MiMaxx.TiO}. Experimental results are taken from
\protect\citeasnoun{91ClElAr.diatom}, \protect\citeasnoun{98LoSiWa.diatom}, \protect\citeasnoun{01LuVexx.diatom} and \protect\citeasnoun{02JeLuVe.diatom}.}
\end{figure}

Figure~\ref{fig:DE} provides estimates of typical experimental and theoretical errors. This figure shows experimental
dissociation energies (crosses) with error bars for early transition
metal oxides, as well as theoretical predictions using a variety of
high-level theoretical methods, specifically icMRCI, icMRCI+Q and
R-CCSD(T) with and without semi-core correlation, all using cc-pVQZ
basis sets. It is clear that the error in theoretical calculations and
experiments are similar, around 500 - 2000 \cm{} from a total
dissociation energy of 45~000 to 60~000 \cm.

\begin{figure}
\includegraphics[width=\textwidth]{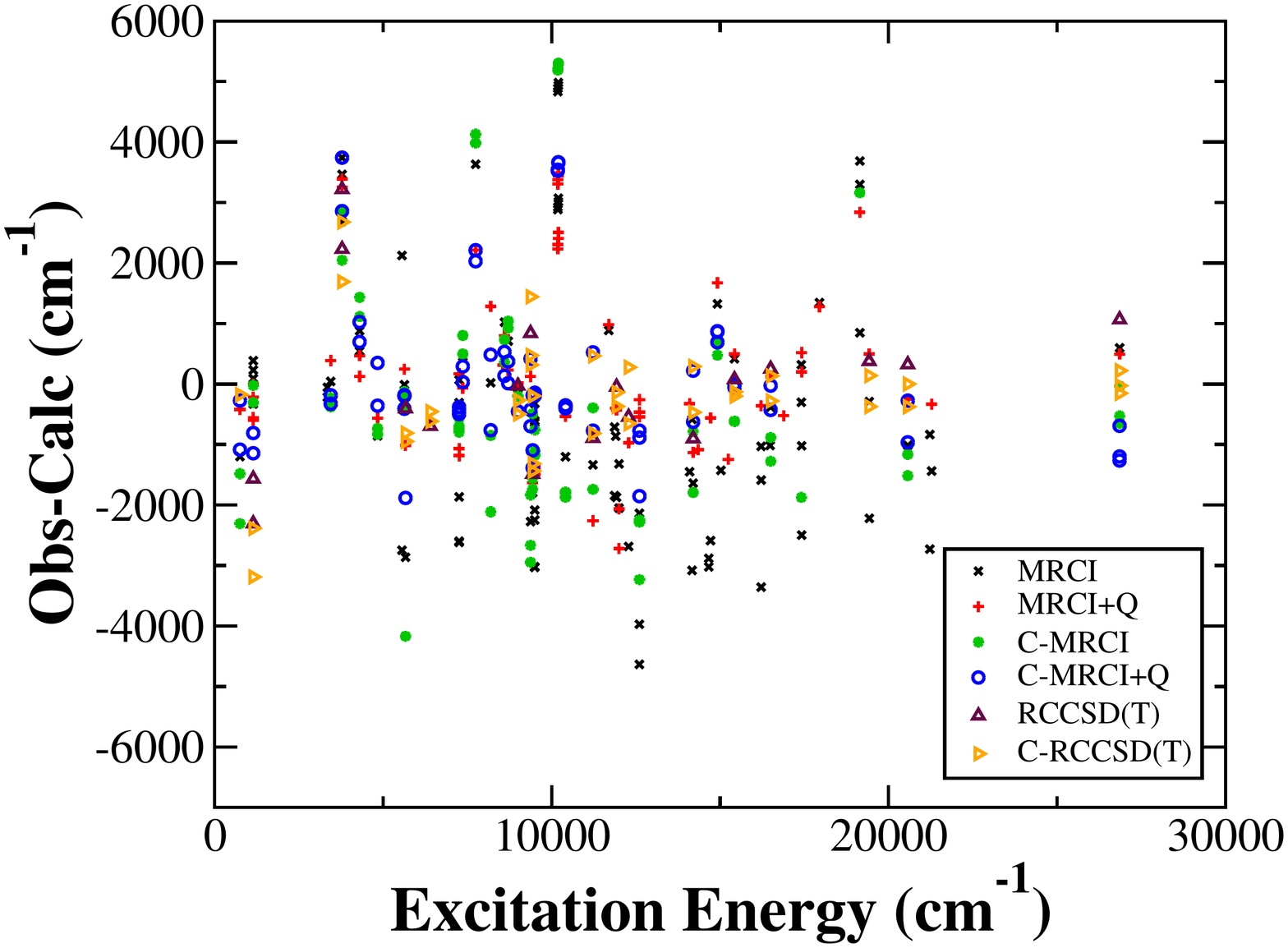}
  \caption{\label{fig:excitationenergy} Observed minus calculated excitation
energies
  of neutral and charged transition metal
    oxides, using large basis sets and a variety of treatments of
    static and dynamic correlation. {\it Ab initio} results taken from
\protect\citeasnoun{88JeKoxx.ScH}, \protect\citeasnoun{97Laxxxx.TiO}, \protect\citeasnoun{01Doxxxx.TiO},
\protect\citeasnoun{07MiMaxx.VO}, \protect\citeasnoun{10MiMaxx.TiO}, \protect\citeasnoun{15HuHoHi.VO}
and \protect\citeasnoun{jt623}.}
\end{figure}

\paragraph*{Electronic excitation energies: }
Even for high quality calculations, the errors in the excitation energies for diatomic molecules often
exceeds 1000 \cm{}, especially for the
higher lying states, as shown in Fig.~\ref{fig:excitationenergy} for
transition metal oxides. The very highest accuracy calculations even
for the lowest electronic states is rarely more accurate than 100 \cm.

\paragraph*{Vibrational Frequency: }

\begin{figure}
\includegraphics[width=\textwidth]{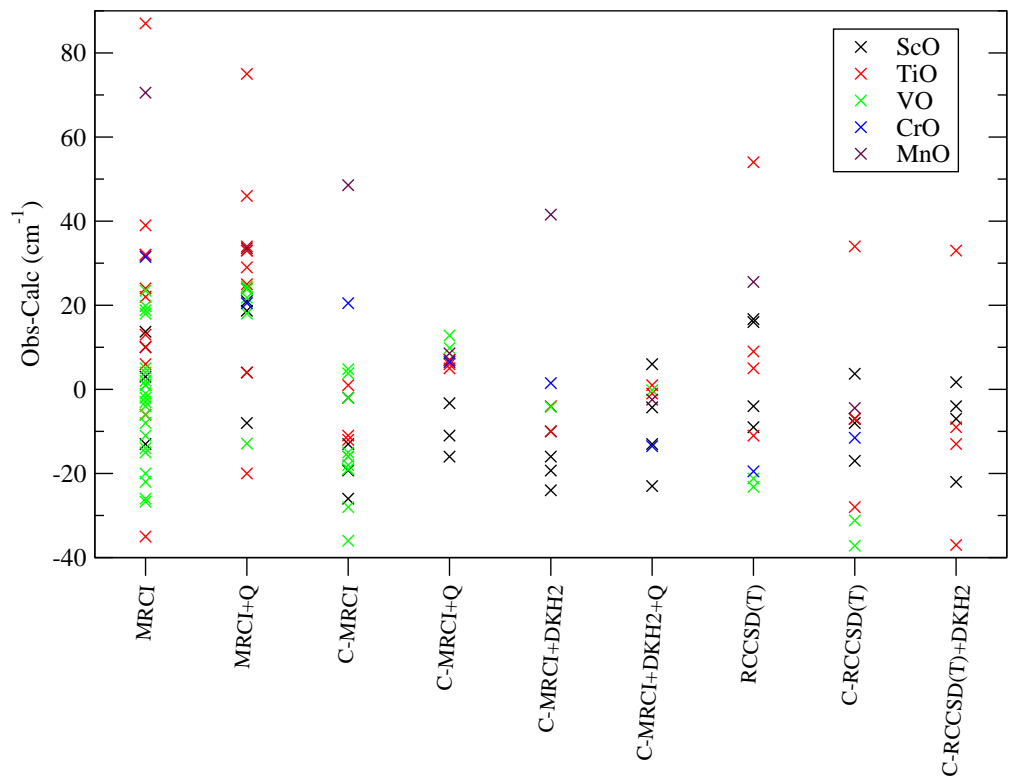}
\caption{\label{fig:vibfreq}  Observed minus calculated in the calculated
vibrational frequencies of
  neutral transition metal oxides, using large basis sets and a
  variety of treatments of static and dynamic correlation. {\it Ab initio} results
  taken from \protect\citeasnoun{97Laxxxx.TiO}, \protect\citeasnoun{07MiMaxx.VO},
\protect\citeasnoun{10MiMaxx.TiO}, \protect\citeasnoun{15HuHoHi.VO} and \protect\citeasnoun{jt623}.}
\end{figure}

Typically, errors in vibrational frequencies are of order 10 to 30 \cm{}, as shown in Fig.~\ref{fig:vibfreq}.
The error seems relatively independent of treatment of dynamic correlation, use of the Davidson correction (+Q),
inclusion of core correlation and relativistic effects.

\paragraph*{Rotational Frequency: }
The equilibrium bond length for each vibronic state determines the
rotational frequency of that state. Generally theory is able to
reproduce experiment quite well. Errors of larger than 0.03 \AA{} for high level
calculations are rare, as shown in Table \ref{tab:Rot}. This table shows
that the errors are normally significantly reduced by the inclusion of
semi-core correlation (i.e. the $3s$ and $3p$ orbitals in the correlated
space).  

A 0.03 \AA{} error in the bond length of a transition metal oxide will
result in an error on order of 0.02 \cm{} for the rotational constant.
This will only cause errors greater than 1 \cm{} for $J>20$. A bond
length error of 0.005 \AA{} will not cause errors in excess of 1 \cm{}
until approximately $J=120$.

\begin{table}
  \caption{\label{tab:Rot} The absolute error in the predicted equilibrium bond distance,
$\Delta r_e$, for various electronic states of TiO and VO.
 The units of bond length are \AA{}.
For TiO, the source of {\it ab initio} data is
\protect\citeasnoun{97Laxxxx.TiO}, \protect\citeasnoun{01Doxxxx.TiO}
and \protect\citeasnoun{10MiMaxx.TiO}.
For VO, the source of {\it ab initio} data is
\protect\citeasnoun{07MiMaxx.VO}, \protect\citeasnoun{15HuHoHi.VO} and \protect\citeasnoun{jt623}.
All {\it ab initio} results included in the averaging are at least icMRCI or
CCSD level,
 with basis sets of triple-zeta quality or better. CAS orbitals are optimised in
different ways. The active space is ($4s$,$4p_x$,$4p_y$,$3d$)/Ti, ($4s$,$3d$)/V
and  ($2p$)/O or ($2s$,$2p$)/O. }
\begin{tabular}{lllccccc}
\hline
& & & \multicolumn{2}{c}{No semi-core} & \multicolumn{2}{c}{With Semi-core} \\
Molecule & State &  Config. &  $\overline{\Delta r_e}$ & $\text{max}(\Delta r_e)$ &  $\overline{\Delta r_e}$ & $\text{max}(\Delta r_e)$  \\
\hline
TiO & X $^3\Delta$ & $4s^13d^1$ & 0.014 & 0.017 & 0.002 & 0.003 \\%
& a$^1\Delta$   & $4s^13d^1$ & 0.014 & 0.018 & 0.004 & 0.005\\
& d$^1\Sigma^+$  & $4s^2$ & 0.011 & 0.017 & 0.005 & 0.006  \\
& E$^3\Pi$  & $4s^13d^1$& 0.022 & 0.027 & 0.002 & 0.003 \\
& A$^3\Phi$ & $4s^13d^1$ & 0.008 & 0.015 & - & - \\
& b$^1\Pi$   & $4s^13d^1$ & 0.014 & 0.022 & - & -  \\
& B$^3\Pi$  &$3d^2$ &0.009 & 0.012 & -  & - \\
& C$^3\Delta$  & $4s^13d^1$ & 0.020 & 0.028 &- & - \\
& c$^1\Phi$ & $3d^2$ & 0.006 & 0.013 & - & - \\
& f$^1\Delta$ & $4s^13d^1$ & 0.008 & 0.017& - & -  \\
VO & X $^4\Sigma^-$  & $4s^13d^2$  & 0.005 & 0.004 & 0.003 & 0.002 \\
& A$^\prime$ $^4\Phi$  & $4s^13d^2$ & 0.004 & 0.004 & 0.004 & 0.003 \\
& c $^2\Delta$ & $4s^23d^1$ & 0.008 & 0.015 & 0.004 & 0.015  \\
& A $^4\Pi$  & $4s^13d^2$ &  0.001 & 0.001 & 0.004 & 0.003 \\
& d $^2\Sigma+$ & $4s^13d^2$ & 0.005 & 0.006 & 0.001 & 0.002  \\
& B $^4\Pi$ & $3d^3$ &  0.005 & 0.005 & 0.010 & 0.005 \\
& e $^2\Phi$   & $4s^13d^2$ & 0.017 & 0.030 & 0.014 & 0.020  \\
\hline
\end{tabular}
\end{table}

\paragraph*{Couplings}

Diagonal spin-orbit coupling removes the degeneracy in energy between
different spin components of non-singlet and non-$\Sigma$ electronic
states.  For $3d$ transition metal diatomics, this is normally treated
as a modification of the pre-existing potential energy curves for an
electronic state; it is generally of magnitude  10 to 250
\cm{}. 
Off-diagonal spin-orbit coupling gives intensity to spin-flipping transitions. 

{\it Ab initio} spin-orbit matrix elements are surprisingly
insensitive to the quality of the calculation.  \citeasnoun{jt599}
investigated the effects of different basis sets and treatment of
dynamic electronic correlation against known experimental results and
found that CASSCF/aug-cc-pVDZ was usually as reliable as the more
expensive methodologies.

Electronic angular momentum matrix elements cannot generally be
derived from experiment and must be calculated {\it ab initio}. Since
there appear to be  no benchmark results, we generally use high level theory, e.g.
icMRCI/aug-cc-pVQZ. The bra and ket wavefunctions do not have to use the
same set of CASSCF orbitals. Thus, we recommend optimising each
separately (for a minimal number of states).  Generally, we find these
matrix elements to be of order 1. 

 Both the spin-orbit and electronic angular
momentum matrix elements have a phase and sign (e.g. can be imaginary)
depending on the phase of the component wavefunctions. These phases need to
be treated carefully \cite{jt589,jt609}.

\subsection{Dipole moments}\label{sec:dipoles}

While it is possible to improve potential energy curves using experimental data, see next section, this
is rarely possible and not generally recommended for dipole moment curves \cite{jt156}.

Although the option of computing dipoles is standard in electronic
structure programs, getting stable results for open shell systems is
far from straightforward. Indeed the literature is full of
examples of dipole moment curves which show unphysical features, see
\citeasnoun{jt573}.  These features can arise from a variety of
causes including (i) using inappropriate models such as Hartree-Fock
or coupled cluster approaches which do not represent the dissociating
wavefunction correctly and  result in large, unphysical dipoles at
large internuclear separations, see \citeasnoun{jt563} for example;
(ii) changes in orbital ordering: our preferred model for computing
dipole moments is icMRCI; in this approach the orbitals are divided into
frozen (fully occupied core orbitals), valence (used in the complete
active and reference spaces) and virtuals (used in the CI step).
Orbital swaps between these spaces at a particular geometry can result
in sharp changes in dipole moments, see \citeasnoun{jt599} for example;
(iii) even if the orbitals do not actually swap they can change
character, for example from ionic to covalent, as a function of bond
length with consequent changes in the dipoles. These changes in
character may well be physical: most ionically bonded species
dissociate to neutral fragments. However, orbital spaces which only contain
orbitals with one of these characteristics can be or become
unbalanced. In this context it should be noted that improved 
transition dipoles moments are obtained if different, optimal orbitals
sets, so called bi-orthogonal orbitals, are used for the two states
involved in the calculation \cite{07ScBaxx.C2}; (iv) symmetry
contamination: as discussed above diatomic species contain significant
symmetry properties, but most quantum chemistry codes do not use
full ($C_{\infty v}$ or $D_{\infty h}$) linear molecule symmetry;
an exception is the package Turbomole \cite{Furche2014}, which however does not
implement multireference methods and is therefore of limited interest to us.

Experience shows that care is needed to obtain smooth dipole moment and coupling curves. We have found
that starting calculations using wavefunctions from a neighbouring geometry and starting calculations which
use large basis sets with orbitals from calculations with smaller basis sets helps to maintain but does
not guarantee smoothness.

Finally we note that measurements of transition absolute transition intensities for open shell systems
and difficult
are rare. An alternative is to measure the lifetime of the excited states.
Such measurements have the significant advantage that they do not require a knowledge of the population
of states in the sample being studied. Lifetimes can be computed routinely
from sets of Einstein A coefficients. This has been done recently for the line lists computed
as part the ExoMol project \cite{jt624}.

\section{Nuclear Motion Calculation and Improving the Model}

The above two sections describe the necessary ingredients for constructing a full {\it ab initio}
spectroscopic model for open shell diatomic species. The resulting nuclear
motion Schr{\"o}dinger equations can then be solved. In the case of {\sc duo} \cite{jt609}
this is done variationally by solving a $J=0$ problem for each, uncoupled PEC and using these
results to provide a basis set to solve the fully-coupled, $J$-dependent problem.
Solutions based on the direct integration of the coupled equations are also possible \cite{94Hutson}.

This direct method works well but is not without some issues. As the number of electronic states
considerd grows, the number of coupling curves that needs to be considered grows as approximately
the square of this number. The generation and handling of these curves can become quite
cumbersome. However there are two further issues. First couplings may occur to states which
are repulsive: this is actually the mechanism for the well-studied phenonomenon of pre-dissociation.
Second, couplings to high-lying states which are not included in the model may also be important.
Both of these situations are probably best dealt with by reverting to a pertubative treatment of
the coupling via the use of effective constants.

Of course, as a model is extended to higher energies, the number of electronic states that
might need to be considered grows rapidly. The issues here is not only the actual
density of states but also the correct treatment of both so-called dark states, ones which
do not have dipole-allowed transitions to the ground state, and the interaction of
quasi-degenerate levels, often called resonances. Such resonances are well-known experimentally
and often lead to intensity stealing by dark states which allows a few levels
associated with these states to be observed. Attributing the correct vibrational (and sometimes even
electronic) quantum numbers the states is often difficult on the basis of observations
alone. For molecules such as C$_2$ the {\it ab initio} theory is now accurate enough for
electronic and vibrational states to be unambiguously identified \cite{15BoMaGo}; this is not
true for systems containing transition metals \cite{jt618}.

One way of improving on the 'direct', {\it ab initio} method for solving the spectroscopic problem is
by addressing  the corresponding inverse problem
\cite{Karkowski2009,Weymuth2014}, that is the task of determining the
potential $V(R)$ which leads to a given set of energy levels $E_{\upsilon
J}$, typically obtained from experiment. One of the oldest ways to perform
this task approximately is to use the semi-classical Rydberg-Klein-Rees (RKR)
method, see \citeasnoun{Karkowski2009}.  A more precise strategy called inverse
perturbation analysis (IPA) based on perturbation theory has been suggested
\cite{Kosman1975,Weymuth2014} and a program implementing
this approach was presented by  \citeasnoun{Pashov2000}.

The gold-standard inverse problem usually involves determining a
potential energy function, usually one represented in some
pre-determined but flexible form, to a set of experimental-derived
data such as energy levels or transition frequencies.  The process
involves minimizing the standard deviation of observed minus
calculated for these data. This methodology is increasing being used
in problems involving coupled potential energy curves, see
\citeasnoun{13YuNiSe.diatom}, \citeasnoun{15WaSeLe.NaH} and
\citeasnoun{jt589}.  A major advantage of performing such fits is
that, at least with the BO approximation, one fit can be used to
obtain spectra for all isotopologues. This means, for example,
that a high accuracy fit to the (plentiful) data available for the
main isotopologue can give excellent predicted spectra for
isotopically substituted species. Of course one can also use data on
several isotopologues simultaneously to determine corrections to the
BO approximation; however, as shown by \citeasnoun{04Watson}, one
needs to be careful here as energy levels alone do not contain enough
information to fully determine all the terms in this situation.

The usual problem in this approach is the lack of the experimental data.
A standard situation is that the number of parameters required
for adequate description of a PEC is greater than the number of vibrational
states present in the experimental set, which makes the inverse problem
under-determined.
In such situations the following approaches are commonly used: (i) approximating
PECs with
simpler model with smaller number of parameters; (ii) fixing some of the
parameters to their
 {\it ab initio} values; (iii) constraining either the PEC or the expansion
 parameters to the {\it ab initio} values; (iv) `morphing' the {\it ab initio} PEC
using a simpler model \cite{jt589}.

Another important issue is to decide on the fitting weights. The common practice
is to set them inverse proportional
to experimental uncertainty \cite{98LeRoy.methods}. A very powerful alternative
is the robust weighting procedure
suggested by \citeasnoun{03Watson.methods}, where the weights are adjusted
dynamically based on the quality of the experimental data.

For systems involving several electronic states using experimental
data to determine the potential also brings the various coupling terms
into play. These too can usually not be fully determined on the basis
of energy levels alone. There is a further problem in  that
fitting can only involve a finite, usually small, number of states but
the couplings link these to higher states not included in the fits.
This issue can be resolved in a number of ways such including
pseudo-states to couple with or representing some of the associated
splittings using effective Hamiltonian parameters. More work is
probably required to fully understand how best to treat this issue.

\section{Conclusion}

In this topical review we address the issue of first principles theoretical
treatment of diatomic molecules with special emphasis on the vibronic spectra
of open shell species containing a transition metal atom.
We advocate the direct solution of the nuclear motion Schr{\"o}dinger equation;
this methodology
is sufficient to obtain quantitative answers. However, the current state-of-the-art
in electronic structure methodologies is not. Indeed the predictions of what
are considered high-level {\it ab initio} electronic structure calculations
are so poor that even the correct ordering of the electronic states is hard to
predict in many cases let alone the precise details required for spectroscopy.

Given this situation, experimental data will be required for the foreseeable
future to help tune spectroscopic models for transition metal containing systems.
In this context, the measurement of a large number of vibronic
bands, rather than very high precision analysis of the rotational
structure of a smaller number of bands is particularly useful. This can be best achieved in
experimental conditions that lead to rotationally cold but vibrational
hot transitions. Similarly, there are rather few absolute intensity measurements
for open-shell transition-metal-containing systems, however radiative lifetime measurements
are equally useful for testing computed Einstein-A coefficients and appear to easier
to perform. New lifetime measurements for a range of systems would be very helpful.

Finally, if anyone is still under the illusion that {\it ab initio} description of
transition metal-containing diatomics is
simple, we suggest that they look at the
potential energy curves of FeO as given in Fig.~1 of \citeasnoun{11SaMiMa.FeO}). This
system, which is of astrophysical importance,
has 48 electronic states below 27000~cm$^{-1}$, even before taking the splitting
of levels into multiple terms by
spin-orbit coupling into account!

\section*{Acknowledgements}
We thank members of the ExoMol team, and in particular Maire Gorman, for many helpful discussions.
This work was supported by the ERC under the
Advanced Investigator Project 267219.
\section*{References}
\bibliographystyle{jphysicsB}

\end{document}